\documentclass[12pt]{article}
\usepackage{amsmath,amssymb,exscale}
\input epsf

\def\gappeq{\mathrel{\rlap {\raise.5ex\hbox{$>$}}
{\lower.5ex\hbox{$\sim$}}}}
\def\permil{$\%\raise.20ex\hbox{$_0$}}
\def\lappeq{\mathrel{\rlap{\raise.5ex\hbox{$<$}}
{\lower.5ex\hbox{$\sim$}}}}

\begin{document}
\topmargin -1.0cm
\oddsidemargin -0.8cm
\evensidemargin -0.8cm
\pagestyle{empty}
\begin{flushright}
UNIL-IPT-00-18\\
IC/2000/133\\
hep-th/0008087\\
August 2000
\end{flushright}
\vspace*{5mm}

\begin{center}
{\Large\bf On some new warped brane world solutions in higher
dimensions}

\vspace{1.0cm}

{\large Seif Randjbar-Daemi$^a$  and
Mikhail Shaposhnikov$^b$}\\
\vspace{.6cm}
{\it {$^{a}$International Center for Theoretical Physics, Trieste,
Italy}}\\
{\it {$^{b}$Institute of Theoretical Physics, University of Lausanne,
\\
CH-1015 Lausanne, Switzerland}}\\

\vspace{.4cm}
\end{center}

\vspace{1cm}
\begin{abstract}

We present new  solutions of higher dimensional Einstein's equations
with a cosmological constant that localize gravity on branes which
are transverse to Ricci-flat manifolds or to homogeneous spaces with
topologically non-trivial solutions of gauge field equations. These
solutions are relevant for the localization of chiral fermions on a
brane.

\end{abstract}

\vfill

\eject
\pagestyle{empty}
\setcounter{page}{1}
\setcounter{footnote}{0}
\pagestyle{plain}


Although the non-compact internal spaces ~\cite{RS1}-\cite{akama}
have been considered in the past, the recent interest, inspired by
ideas from  superstring theory, emphasizes the notion that our
large-scale universe may be  regarded as a brane embedded in a
higher-dimensional manifold.  The gravity may be localized on a brane
by the Randall-Sundrum mechanism \cite{rusu}, while the observed
particles are attached to a brane because of dynamics of string
theory \cite{branes} or by field-theoretic effects \cite{RS1},
\cite{Dva}-\cite{rash}. It thus becomes relevant to look for  general
solutions of the $D$-dimensional field equations. It is expected
that  a richer geometrical and topological structure of the subspaces
spanned by  extra dimensions will lead to more interesting physical
consequences on the brane world. A number of warped solutions have
already been found in \cite{RS2,seif}, \cite{ck}-\cite{cehs}.

In this note, we consider a $D = D_1 + D_2 + 1$-dimensional system of
gravity - Yang-Mills system and look for warped solutions of the
form
\begin{equation}
\label{feq2}
{\rm d}s^2={\rm e}^{A(r)}\eta_{\mu\nu}{\rm d}x^\mu{\rm d}x^\nu
+{\rm e}^{B(r)}g_{mn}(y){\rm d}y^m{\rm d}y^n+{\rm d}r^2,
\end{equation}
where $x^0,x^1,\dots,x^{D_1-1}$ cover the brane world $M_{D_1}$ and
$y^1,\dots,y^{D_2}$ cover an internal space $K$ with a metric
$g_{mn}(y)$. The  metric with $D_2=0$ and $A=c|r|,~c < 0$ is the
Randall-Sundrum case~\cite{rusu}, the one with $D_2=1$ and $A = B =
c|r|,~c < 0$ is the case of local string \cite{gs}, $D_2=1$ and $A =
c|r|,~c < 0,~ B=const$ corresponds to a global string \cite{rg}.
When  $D_2 \geq 2$ the approximate solutions with scalar
fields~\cite{ov} or exact bulk  solutions with $p$-form
fields~\cite{hedg} have all assumed that $K=S^{D_2}$.  In this note
we shall consider cases where $K$ is not restricted to being a
sphere. These general cases become important if we want to localize
chiral  fermions to the brane world $M_{D_1}$~\cite{rash}.

In the presence of Yang-Mills fields we shall assume that $K$ is a
symmetric  homogeneous space $G/H$ with a $G$-invariant metric
$g_{mn}(y)$ defined on it.  We shall also assume that the gauge group
has a non-trivial intersection with  $H$. For example, for $G/H=S^4$
it is sufficient to assume that the gauge  group contains an SU(2)
subgroup. On such spaces one can construct  $G$-invariant solutions
to the Yang-Mills equations~\cite{Rand}, giving
\begin{equation}
\label{eq2}
\vec{F}_{mp}\cdot\vec{F}_n^{\,p}={{\rm e}^{-B}\over D_2}F^2g_{mn},
\end{equation}
where, by virtue of $G$-invariance,
\begin{equation}\label{eq3}
F^2\equiv g^{mp}g^{nq}\vec{F}_{mn}\cdot\vec{F}_{pq}
\end{equation}
must be a constant. The non-trivial equations to be solved can then
be put  into the form
\begin{equation}
A''+{D_1\over 2}{A'}^2+{D_2\over 2}A'B'= {4\kappa^2\over D-
2}\left(-\Lambda+{F^2\over 4g^2}{\rm e}^{-2B}\right)~,
\label{eq4}
\end{equation}
\begin{equation}
B''+{D_2\over 2}{B'}^2+{D_1\over 2}A'B'= {4\kappa^2\over D-
2}\left(-\Lambda-{2D-D_1-4\over D_2}{F^2\over 4g^2}{\rm e}^{-2B}\right)
+k{\rm e}^{-B}~,
\label{eq5}
\end{equation}
\begin{equation}
D_1A''+D_2B''+{D_1\over 2}{A'}^2+{D_2\over 2}{B'}^2= {4\kappa^2\over D-
2}\left(-\Lambda+{F^2\over 4g^2}{\rm e}^{-2B}
\right)~,
\label{eq6}
\end{equation}
where $\Lambda$ and $\kappa^2$ are $D$-dimensional cosmological and
Newtonian constants respectively, $g$ is the gauge coupling. Also the
constant $k$ in eq. (\ref{eq5}) is the curvature scalar
of $K$ defined by $R_{mn}=kg_{mn}$. It can be  shown that only two
out of the three equations above are independent. It  should also be
noted that these equations are valid in the bulk, the brane core region
around $r=0$ should be treated in the same way as in~\cite{gs,hedg},
which provides a relationship between different tensions on the brane
and the values of $A'$ and $B'$ near the origin.

At this point it  is convenient to consider the cases $k=0$ and
$k\neq 0$ separately.

{\bf (i) $k=0$.} In this case $K$ is a Ricci flat manifold. Well known
examples are tori and Calabi-Yau manifolds.
Setting $F^2=0$, one can find all solutions of eqs.
(\ref{eq4}-\ref{eq6}), generalizing the solutions found in
\cite{RS2,seif,hedg} for $D_1=4,~ D=6$, where the $k$-term is
trivially absent. The general result is:
\begin{equation}
A(r) = a\, log [z'(r)] + b\, log [z(r)],~~B(r) = c\, log [z'(r)] + d\, log
[z(r)]
\end{equation}
where
\begin{equation}
a=\frac{2}{D-1}\left( 1-\sqrt{\frac{D_2(D-2)}{D_1}}\right)~,~~
b=\frac{2}{D-1}\left( 1+\sqrt{\frac{D_2(D-2)}{D_1}}\right)~,
\end{equation}
\begin{equation}
c=\frac{2}{D-1}\left(1+\sqrt{\frac{D_1(D-2)}{D_2}}\right)~,~~
d=\frac{2}{D-1}\left(1-\sqrt{\frac{D_1(D-2)}{D_2}}\right)~,
\end{equation}
and
\begin{equation}
z(r) = Re\left(\alpha e^{\gamma r} + \beta e^{-\gamma r}\right),~~
 \gamma = \sqrt{-{(D-1)\kappa^2\Lambda\over 2(D-2)}}~,
 \end{equation}
 where $\alpha$ and $\beta$ are arbitrary constants.

It is easy to  verify that for a negative cosmological constant there
is a simple solution  with $A = B = -\gamma r$, where $\gamma$ is not given by (10) but $\gamma =
 \sqrt{-{8\kappa^2\Lambda\over (D-1)(D-2)}}$ . As in the standard RS
solution, we  can extend the range of $r$ to $(-\infty,+\infty)$ and
write the solution as  $A = B = -\gamma |r|$. In this case $A''=B''= - 2
\gamma\delta(r)$. The field equations in the core  region will then relate
the tension of the brane to $\gamma$, along the lines  discussed
in~\cite{rusu,gs,hedg}. At infinity, $r \rightarrow \infty$, the
square of the curvature tensor is singular, as follows from the
general expression, given in \cite{seif}. The only exception is when the
internal space is flat, for example, it can be a torus. In this case the soltuion is
non singular even as $r \rightarrow \infty$. For the general solution to make
physical sense the singularity should be smoothen by, say, string
theory effects.

{\bf (ii) $k\neq 0$.} In this case the $G$-invariant solutions of
Yang-Mills  equations can be constructed as in \cite{Ra}. One can
have a regular solution in the bulk with the structure $A{\rm
d}S_{D_1+1}\times G/H$  with $A=-c r$ and $B$ a constant, similar to
the one considered in \cite{hedg} for a compactification on a monopole
for $D_2=2$. One can consider two possibilities. In the first case we
can treat $r$ as a radial co-ordinate, varying fom zero to infinity.
Then this  solution is singular at $r=0$ and requires the addition of
a $G$ invariant brane, residing at this point.  The tensions on the
brane must then be fine tuned accordingly. In the second case $r$ may
vary from $-\infty$ to $+\infty$. As the Einstein equations are
symmetric with respect to the transformation $r\rightarrow -r$, a
solution, leading to the localization of gravity, looks like $A=-c
|r|$ and requires the presence of a brane as well.

An interesting question arises whether one can construct a braneless
solution that is regular at $r=0$  and which approaches the bulk
solution just described for large values of $r$. Arguing  along the
lines of ref. \cite{seif}, this seems to be impossible, at least if
$-\infty < r < \infty$, when the $Z_2$ symmetry is imposed. To this
end it is convenient to rewrite eqs.(\ref{eq4}-\ref{eq6}) as a single
equation for $B$ only,
\begin{equation}
B''=\frac{1}{2\,
    (D-2 ) \,(D_1 -1 ) \,D_2\,
    e^{2\,B}}
     \left\{( D -2 ) \,D_2^2\,{B'}^2\,e^{2\,B}\,
      +\right.
     \end{equation}
\[
    \left. + 2\,( D_1 -1 ) \,
     \left( ( 4 - 2\,D + D_1)\,\frac{\kappa^2 \,F^2}{g^2}-
       D_2\,
        \left( 2\,k - D\,k + 4\,e^B\,{\kappa}^2\,\Lambda \right)\,e^B \right)
    \pm S \right\}
\]
where
\begin{equation}
S=D_2\,\sqrt{D_1(D -2 )}\,
  B'\,e^B\,\left\{D_2\,(D -2)^2 \,
       {B'}^2\,e^{2\,B} \right. -
       \label{S}
      \end{equation}
\[
     \left.  2\,(D_1 -1) \,
        \left( (2\,D - 2\, D_1 -3 ) \,
              \frac{\kappa^2 \,F^2}{g^2}+ 4\,(D_1 -1 )
         \,e^{2\,B}\,\kappa^2\,\Lambda
              + D_2\,
          \left( ( 2 - D ) \,k +
      4\,e^B\,\kappa^2\,\Lambda \right)\,e^B
         \right)\right\}^{\frac{1}{2}}
\]
Now, this equation describes a motion of a particle in a potential
$V(B)$
which is obtained by setting $B'=0$ in eq. (\ref{S}) and integrating
with respect to $B$.
\begin{equation}
V(B)= k\,e^{-B} - \frac{1}{\left(D-2\right)}
\left\{ \frac{1}{2 D_2} \left(2\,D - D_1 -4 \right)
\,\frac{\kappa^2\,F^2}{g^2}\,e^{-2\,B}
 - 4\,{\kappa}^2\,\Lambda\, B\right\}
 \label{pot}
\end{equation}
with quite complicated friction force. The regularity of solution at
$r=0$ requires $B'(0)$=0, while  $B \rightarrow const$ at
infinity.  In a mechanical analogy the particle should start moving
with zero velocity and reach the maximum of the potential at ``time"
$r \rightarrow \infty$. This is only possible if the potential
(\ref{pot}) has two extrema. This requirement leads to a certain
constraints on the parameters   $\frac{\kappa^2 \,F^2}{k g^2}$,
$\frac{\kappa^2\,\Lambda}{k}$ and on the dimensionalities $D_1$ and
$D_2$. The latter happen to be not consistent with reality of  the
quantity $S$ defined in eq. (\ref{S}) at the maximum of the
potential.

Both types of solutions (i) and (ii) can be used to obtain localized
chiral fermions in  $M_{D_1}$. As argued in~\cite{rash} for the case
$K=K_3$ and $D=9$, the index  of the Dirac operator
$\not\hspace{-2mm}\nabla$ acting in the transverse space is two,
which implies the existence of two families of  chiral fermions in
$M_4$. Alternatively, if we choose $K=S^4$ and $\vec{F}$ an  $SU(2)$
instanton configuration on $S^4$ then we obtain ${2\over
3}t(t+1)(2t+1)$ generations of chiral fermions on $M_4$, where $t$ is
the  $SU(2)$ spin of the fermions \cite{Ra}.

In conclusion, we found new solutions that localize gravity on a
brane with non-trivial transverse spaces and gauge instanton
backgrounds. It remains to be seen if a realistic theory
incorporating fields of the Standard Model can be constructed on
these solutions.

We wish to thank T. Gherghetta for helpful discussions. S. R.-D. is
thankful to IPT-UNIL for hospitality. This work was supported by the
FNRS,  contract no. 21-55560.98.

\end{document}